\documentclass{article}
\usepackage{graphicx}
\usepackage{subcaption}
\usepackage{float}
\usepackage{geometry}
\usepackage{authblk}
\usepackage{url}
\usepackage{amsmath}
\usepackage{booktabs} 
\usepackage{multirow}
\usepackage{hyperref}
\usepackage{float}
\usepackage{comment}
\usepackage[round]{natbib}

 \hypersetup{
    colorlinks=true,
    linkcolor=black,
    citecolor=black,
    urlcolor=black
}

\begin{document}

\title{A novel density-based approach for estimating unknown means,  distribution visualisations and meta-analyses of quantiles}

\author[1,$\ast$, $\dag$]{Alysha M De Livera}
\author[1, $\dag$]{Luke Prendergast}
\author[1]{Udara Kumaranathunga}

\affil[1]{Mathematics and Statistics, School of Computing, Engineering and Mathematical Sciences, La Trobe University, Kingsbury Dr, 3086, VIC, Australia }

\date{}

\maketitle
\vspace{-1.2cm}
\begin{center}
    $\ast$ Corresponding author: {Alysha M De Livera} 
 \url{Email: a.delivera@latrobe.edu.au}\\
        $\dag$ Equal contribution\\
    \date{\today}
\end{center}

\section*{Abstract}

In meta-analysis with continuous outcomes, the use of effect sizes based on the means is the most common. It is often found, however, that only the quantile summary measures are reported in some studies, and in certain scenarios, a meta-analysis of the quantiles themselves are of interest. We propose a novel density-based approach to support the implementation of a comprehensive meta-analysis, when only the quantile summary measures are reported. The proposed approach uses flexible quantile-based distributions and percentile matching to estimate the unknown parameters without making any prior assumptions about the underlying distributions. Using simulated and real data, we show that the proposed novel density-based approach works as well as or better than the widely-used methods in estimating the means using quantile summaries without assuming a distribution apriori, and provides a novel tool for distribution visualisations. In addition to this, we introduce quantile-based meta-analysis methods for situations where a comparison of quantiles between groups themselves are of interest and found to be more suitable. Using both real and simulated data, we also demonstrate the applicability of these quantile-based methods.

\textbf{Keywords} meta analysis, quantiles, estimating means, distribution visualisation

\section{Introduction}\label{sec1}

Perhaps the most popular reference text for meta-analysis is by \cite{borenstein2011introduction} and consequently we will employ similar notation.   Let $Y_1,\ldots,Y_K$ denote statistical estimators of a population parameter $\theta$, called the \textit{effect}, arising from $K$ independent studies.  We assume that the $Y_k$s are normally distributed (or approximately normally distributed) and let $v_k$s denote the estimator variances; i.e. $\text{Var}(Y_k)=v_k$ $(k=1,\ldots,K)$ so that $Y_k\sim N(\theta,v_k)$.  From this we can form the fixed effects model (FEM),
\begin{equation}
    Y_k=\theta + \epsilon_k,\\ k = 1,\ldots, K\label{FEM}
\end{equation}
where $\epsilon_k$ is the sample error term for the $k$th study with $\epsilon\sim N(0, v_k)$.  The purpose of a FEM meta-analysis is to pool the estimators from the $K$ studies to obtain an overall estimate of the effect, $\theta$.  This is usually as a weighted average,
$$\widehat{\theta} = \sum^K_{k=1}w_k Y_k$$
with weights $w_1,\ldots,w_K$ summing to one.  The choice of weights that lead to the minimum variance of the estimator $\widehat{\theta}$ are $w_k=W_k/\sum^K_{k=1}W_K$ where $W_k=1/v_k$ scaled to sum to one, $k=(1,\ldots K)$.  Since the $W_k$ is inverse of the estimator variances for the $k$th study, studies with smaller estimator variance are given larger weights. This approach is the commonly implemented inverse variance weights (IVW) meta-analysis.

In practice, the FEM is rarely justifiable unless each study has been conducted under the exact same conditions and with the samples that are each representative of the same underlying population.  For example, demographics of study participants (e.g., age, gender, nationality etc.), treatment regimes (e.g., drug dosage, length of treatment etc.) and so on, usually differ between studies.  Intuitively, such differences should result in differences to the effect of interest, $\theta$.  In meta-analysis, we refer to this difference in effects across studies as \textit{heterogeneity}.

One way of accounting for heterogeneity in our modelling is to use a random effect, $\gamma_k$.  This leads to the random effects model (REM),
\begin{equation}
    Y_k=(\theta + \gamma_k) + \epsilon_k,\\ k = 1,\ldots, K\label{REM}
\end{equation}
where $\gamma_k\sim N(0, \tau^2)$ and $\tau^2\geq 0$ is the heterogeneity variance.   From this point, we will use the REM with assumed heterogeneity and simply note that the FEM is a special case of the REM (i.e., $\tau=0$).  The REM IVW now needs weights that include two sources of variation so that $W_k^*=1/(v_k+\tau^2)$.   The variance of the IVW estimator is simply the inverted sum of unscaled weights so that $$\text{Var}(\widehat{\theta}^*)=\left[\sum^K_{k=1}W_k^*\right]^{-1}$$ where $\widehat{\theta}^*$ is the REM estimator of $\theta$ using the $W_k^*$ weights.

Since $\tau^2$ is unknown, it needs to be estimated, and several estimators are commonly found in practice.  The DerSimonian and Laird estimator \citep[DL,][]{dersimonian1986meta} is a simple to compute estimator of $\tau$ and is consequently found in many statistical packages.  However, many other estimators of $\tau$ are available, and we favour the restricted maximum likelihood (REML) estimator that is the default for the \texttt{metafor} package \citep{metafor} available in R \citep{R}.  That said, there is no single estimator of $\tau$ that always outperforms all others \citep{viechtbauer2005bias}, although the REML estimator appears a good choice overall. 

Measures of heterogeneity are important since they help researchers to understand the broader population of effects and to draw valid conclusions.  For example, under the REM in \eqref{REM}, $\theta$ is the average effect and we can use an estimate of $\theta$ to make conclusions about this average effect.  However, when heterogeneity is present (i.e. $\tau>0$) effects across different sub-populations or under different treatment regimes may vary considerably from the average effect.  The most common measure of heterogeneity is $I^2 \in [0,1]$ \citep{higgins2002quantifying} although it is often misinterpreted \citep{borenstein2017basics}.  $I^2$ is a measure of the proportion of variation in the observed effects that is not attributed to the variance of the effect estimators.  Hence, a large $I^2$ indicates a large amount of heterogeneity relative to the total amount of variability in effects.  There has also been a push to using prediction intervals \citep{inthout2016plea} since they will tell researchers, with some degree of confidence, a range that captures most (e.g. 95\%) effects, including effects from future studies.

\subsection{Introduction to the problem}

For continuous data, effect sizes based on means are the most common. The standard meta-analysis approaches described above assume approximate normality, but not in the underlying distribution from which the data is sampled. When the underlying distribution is drawn is skewed or when the sampled data contains outliers, it is common for researchers to report medians and interquartile ranges (or ranges) to summarise their findings.  Hence, estimating the unreported means and standard deviations using only the reported quantile summary measures~\citep{hozo2005, wan2014, bland2015,luo2018,shi2020,mcgrath2020} so that they can be used in a meta-analysis with other studies that have reported means has been an active research area in the recent meta-analysis literature. 

If the distribution from which the study data is drawn is skewed, then the mean and standard deviation can be poor measures of centrality and dispersion.  Additionally, if the data contains outliers, then these outliers can adversely influence moment-based estimators.   In these situations researchers may consider meta-analysis of medians and other quantiles instead of means to draw more meaningful conclusions. Furthermore, in some scenarios, medians and other quantiles are widely accepted summary statistics (e.g., median income, survival time among others), and so, converting these to means and standard deviations does not make sense. In these situations, meta-analysis of quantiles is more intuitive and relevant.  Recent literature has proposed ways of meta analysing medians~\citep{CharlesTiGray2020TaMo, ozturk2020meta} and meta-analysis of quantiles and functions of quantiles are therefore areas of further interest.

Additional to the above, it is important to note that very little about distribution shape and skew can be found using just means and standard deviations.  However, medians, quartiles, ranges etc. provide very rich information in this regard, and can therefore add great value to a meta-analysis.   In the next section, we provide a brief summary of related work that aims to handle the above challenges and our contribution in this paper.

\subsection{Brief summary of related work and our contribution}

We now briefly summarise some recent work that has helped motivate our contributions. 

\subsubsection{Estimating unknown means and standard deviations}\label{subsubsec:est_means_sd}

Several excellent recent studies~\citep{hozo2005, wan2014, bland2015,luo2018,shi2020,mcgrath2020} have proposed approaches for estimating the sample mean and standard deviation from commonly reported quantiles in meta-analysis. These studies have considered the following three scenarios with the sample sizes being available in each case:
\begin{align*}
S_1 &= \{a, m, b\} \\
S_2 &= \{q_1, m, q_3\} \\
S_3 &= \{a, q_1, m, q_3, b\},
\end{align*}
where $\{a, q_1, m, q_3, b\}$ denote the sample minimum, first quartile (i.e., an estimate of the 25\% percentile), the median, third quartile and the maximum respectively. In the widely-used R packages \texttt{metafor} \citep{metafor} and \texttt{meta} \citep{metapackage} that implement meta-analysis estimation, the transformation method proposed by \cite{luo2018} is the default method for estimating the mean in all three scenarios. For estimating the standard deviations, the methods proposed by \cite{wan2014} are the default methods in scenarios $S_1$ and $S_2$, while the approach proposed by \cite{shi2020} is used for scenario $S_3$. As \cite{shi2023detecting} explains, in the presence of skewness, the transformations that are built on the normality assumption for the underlying data and the conclusions from the meta-analysis can be unreliable. 

In this paper, in Section~\ref{sec:estimaing_means}, we introduce a different approach that uses flexible densities to approximate the underlying densities based on the quantile summaries. In Section~\ref{subsec:sim_means}, we then compare the proposed approach with the approaches mentioned above for estimating the means and standard deviations. 

\subsubsection{Distribution estimation and vizualisations}

Forest plots are a very popular visualisation tool used in meta-analysis.  At a minimum, they typically depict confidence intervals for individual studies, weights assigned to each study in the pooled estimator and point and interval estimates of the pooled effect. They therefore provide a convenient summary of the findings from each study that are used in the meta-analysis.  For further details, including for other visualisation tools, see, e.g., \cite{anzures2010graphical}. When only the quantile summaries and sample sizes are available, recently, ~\cite{shi2023detecting} proposed skewness tests as a way of summarizing skewness of data. 

In Section \ref{sec:visual}, we propose a new tool to further summarize important information about the populations within each study, providing further insight into heterogeneity, skewness and differences between groups.   To do this, we use only the quantile summaries to estimate and plot the underlying densities. An application of this tool to a real example involving COVID-19 data is given in Section~\ref{subsec:dist_visualisation_eg} where we also show that visual displays of the densities can provide additional and valuable insights.  

\subsubsection{Meta analysis of quantiles}

The approximate variance of a quantile estimator is proportional to $1/f(x_p)$ where $f$ is the density function from which the data is sampled and $x_p$ is the quantile to be estimated.  Hence, to obtain an of the variance we require an estimate of the density function. Recently, IVW meta-analysis of medians using approximations to the density function have been considered \citep{CharlesTiGray2020TaMo, ozturk2020meta}.  

In Section \ref{sec:ma_quant}, we introduce estimating the density from the reported quantiles so that an approximation to the variance of the median and other quantiles, can be obtained for meta-analysis. While some simulations are presented in this section, we also present an application of these concepts to real data in Section~\ref{subsec:quant_ma_eg}.

\section{Methods}\label{sec:proposed}

In this section we consider estimation of the underlying densities from quantiles, and consider applications of these estimates in the context of meta-analysis.

\subsection{Methods for estimating mean and standard deviations from quantile summaries}\label{sec:estimaing_means}

Consider a random variable $X\sim F$ where $F$ is the distribution function of $X$.  Let $Q$ denote the associated quantile function such that, $x_p = Q(p)$, $p\in [0,1]$ where $P(X\leq x_p)=p$.  We shall refer to $x_p$ as the $p$th quantile.  The quantile function uniquely defines a distribution in that no two different distributions can have the same quantile function.  Hence, the quantile function contains all we need to know about a distribution and can be used to obtain moments of $F$ directly.  In particular, we can obtain the mean and the variance of $X$ respectively using \citep[e.g., p68 and Table 3.1 of][]{gilchrist2000statistical},
\begin{equation}
    E(X)  =\mu = \int^1_0 Q(p)dp\label{mu}
\end{equation}
and
\begin{equation}
    \text{Var}(X)  =\sigma^2 = \int^1_0 \left[Q(p)\right]^2dp - \mu^2\label{sigma2}.
\end{equation}

Hence, if we have an estimate of a quantile function, $\widehat{Q}$, then we can obtain estimates of the mean and variance through integration of functions of $\widehat{Q}$.  Since this integration is bounded over $[0,1]$, this is a straightforward task using numerical integration techniques.  We will use these properties of the quantile function to estimate means and standard deviations when only quantile summary measures are provided.  

\subsubsection{Using a five-point summary 
$S_3 = \{a, q_1, m, q_3, b\}$}\label{sec:5_point_summary_est}

Let $\{a, q_1, m, q_3, b\}$ denote the sample minimum, first quartile (i.e. the estimate of the 25\% percentile), the median, third quartile and the maximum respectively.  The Generalised Lambda Distribution (GLD) is a flexible four parameter family of distributions, which equates to well-known distributions such as uniform, exponential and logistic, and closely approximates several others such as log-normal, Cauchy and normal. GLD, using the FKML parameterisation~\citep{Freimer1988}, is defined by its quantile function 
\begin{equation}
    Q(u; \lambda_1, \lambda_2, \lambda_3, \lambda_4) = \lambda_1 + \frac{1}{\lambda_2} \left( \frac{u^{\lambda_3} - 1}{\lambda_3} - \frac{(1-u)^{\lambda_4} - 1}{\lambda_4} \right) = \lambda_1+\frac{1}{\lambda_2}S(u;\lambda_3,\lambda_4),\label{GLD_Q}
\end{equation} where $\lambda_1$ is a location parameter, $\lambda_2>0$ is an inverse scale parameter, and $\lambda_3$ and $\lambda_4$ are shape parameters. A description of several methods that have been developed for estimating the parameters of the GLD is given by ~\cite{dedduwakumara2019simple}. 

When only the quantile information is available, we can find estimates for the GLD parameters by using a series of simultaneous equations involving the estimated quantiles and equated with the GLD population counterparts such as was done by \cite{karian1999fitting}.  Namely, we define
\begin{equation}
    \widehat{\rho}_1 = m,\; \widehat{\rho}_2 = b - a,\; \widehat{\rho}_3 = (m - a)/(b - a),\; \widehat{\rho}_4=(q_3 - q_1)/\widehat{\rho}_2.\label{rho_hats}
\end{equation}

Note that the sample minimum and maximums, $a$ and $b$, are estimates to the $Q(0.5/n)$ and $Q(1-0.5/n)$ quantiles.  Hence, the $\widehat{\rho}_i$s above are estimates to
\begin{align*}
\rho_1 &= \lambda_1 + \frac{1}{\lambda}_2 S(0.5;\lambda_3,\lambda_4),\\
    \rho_2 &= S(1 - 0.5/n; \lambda_3,\lambda_4) - S(0.5/n; \lambda_3,\lambda_4),\\
    \rho_3 &= \frac{S(0.5; \lambda_3,\lambda_4) - S(0.5/n; \lambda_3,\lambda_4)}{S(1 - 0.5/n; \lambda_3,\lambda_4) - S(0.5; \lambda_3,\lambda_4)}\\
    \rho_4 &= \frac{S(0.75; \lambda_3,\lambda_4) - S(0.25; \lambda_3,\lambda_4)}{\rho_2}
\end{align*}
where both $\rho_3$ and $\rho_4$ depend only on $\lambda_3$ and $\lambda_4$.  While there is no closed form solution for obtaining estimates to $\lambda_3$ and $\lambda_4$ from $\rho_3$ and $\rho_4$ and their estimates, we can use simple optimisation strategies to find solutions to
\begin{equation}
    \{\widehat{\lambda}_3,\widehat{\lambda}_4\}=\underset{\lambda_3,\lambda_4}{\mathrm{argmin}}\left[\left(\widehat{\rho}_3 -\rho_3\right)^2+\left(\widehat{\rho}_4 -\rho_4\right)^2\right].\label{est_l3_l4}
\end{equation}

We can then use the estimates of $\lambda_3$ and $\lambda_4$, together with $\widehat{\rho}_2$ replacing $\rho_2$, to obtain an estimate for $\lambda_2$ as
\begin{equation}
    \widehat{\lambda}_2 = \frac{1}{\widehat{\rho}_2}\left[S(1-0.5/n;\widehat{\lambda}_3,\widehat{\lambda}_4)-S(0.5/n;\widehat{\lambda}_3,\widehat{\lambda}_4)\right]\label{est_l2}
\end{equation}
and similarly for $\lambda_1$,
\begin{equation}
    \widehat{\lambda}_1=\widehat{\rho}_1 - \frac{1}{\widehat{\lambda}_2}S(1/2;\widehat{\lambda}_3,\widehat{\lambda}_4).
\end{equation}
    
In many cases, a study will have two arms, e.g. a treatment arm and a control group.  When this is the case, we could use the single arm estimators as above on each of the groups, or use some shared information across to improve our estimates.  Here, we consider the case where the underlying distribution in each arm has the same shape, and differ only in location and scale.  This would be true with location-scale distributions, but also others if the simple form of the quantile function adequately represents the shape for both.  Here, the GLD approximating distribution would have equal or very similar $\lambda_3$ and $\lambda_4$, with $\lambda_1$ and $\lambda_2$ allowed to vary.  For example,
\begin{description}
    \item[Treatment Group:] $X\sim \text{GLD}(\lambda_{t1},\lambda_{t2},\lambda_3,\lambda_4)$. 
    \item[Control Group:] $Y\sim \text{GLD}(\lambda_{c1},\lambda_{c2},\lambda_3,\lambda_4)$. 
\end{description}

Let $\{a_t, q_{t1}, m_t, q_{t3}, b_t\}$ and $\{a_c, q_{c1}, m_c, q_{c3}, b_c\}$ denote the five-point summaries for the treatment and control arms respectively.  Additionally, let $n_t$ and $n_c$ denote the sample sizes and associated weights $w_t=n_t/(n_t+n_c)$ and $w_c=n_c/(n_t+n_c)$ so that $w_t+w_c=1$.  To allow for different sample sizes between the two groups, we also need to define $\rho_{t3}$ and $\rho_{t4}$ for the treatment arm which are the $\rho_3$ and $\rho_4$ from above but with $u_t = 0.5/n_t$.    Similarly, we define $\rho_{c3}$ and $\rho_{c4}$ for the controls.  Note, when the sample sizes are equal, we have $\rho_{t3} = \rho_{c3}$ and $\rho_{t4} = \rho_{c4}$.

Then a possible estimator for $\lambda_3$ and $\lambda_4$ is
\begin{equation}
    \{\widehat{\lambda}_3,\widehat{\lambda}_4\}=\underset{\lambda_3,\lambda_4}{\mathrm{argmin}}\left[w_t\left(\widehat{\rho}_{t3} -\rho_{t3}\right)^2+w_t\left(\widehat{\rho}_{t4} -\rho_{t4}\right)^2+w_c\left(\widehat{\rho}_{c3} -\rho_{c3}\right)^2+w_c\left(\widehat{\rho}_{c4} -\rho_{c4}\right)^2\right].\label{est_l3_l4_two_arm}
\end{equation}
where $\widehat{\rho}_{t3}$, $\widehat{\rho}_{t4}$, $\widehat{\rho}_{c3}$ and $\widehat{\rho}_{c4}$ are the estimates for $\rho_{t3}$ etc. defined above.  We have used weights here since it is reasonable to assume that better estimates of $\rho_3$ and $\rho_4$ are obtained with larger sample sizes, and this would be particularly true for minimums and maximums as quantile estimates.  Hence, we can put more emphasis on the group with the larger sample size.

\begin{figure}[h!t]
    \centering
    \includegraphics[scale=0.5]{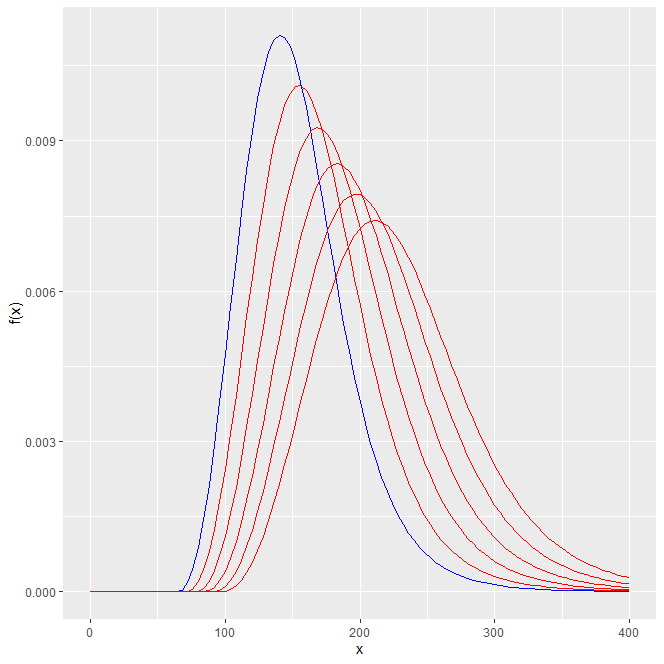}
    \caption{Density plots of $X\sim \text{LN}(5, 0.25)$ (in blue) and then density plots of the $1.1\times X$, $1.2\times X$, $1.3\times X$, $1.4\times X$, $1.5\times X$ random variables (all in red).  These distributions all have the same shape and skew parameters, differing only in location and scale.}
    \label{fig:lognorm_mult}
\end{figure}

Examples under which the GLD remains invariant with respect to the $\lambda_3$ and $\lambda_4$ parameters include, and most obviously, normal distributions for which the GLD approximation differs only in $\lambda_1$ and $\lambda_2$, exponential distributions, and other distributions that do not include skew or shape parameters. However, in practice there can be many more scenarios for which one group differs from another through either a location shift, or a scale factor that effects both location and scale.  For example, consider the $X\sim \text{LN}(5, 0.25)$.  Using the online web application at \url{https://mathstats-drama.shinyapps.io/GLDlambdas/} \citep{dedduwakumara2019simple}, GLD$(146.5231, 0.0399, 0.3055, -0.0412)$ provides an extremely close approximation (Hellinger's distance of 0.0007).  This density is shown in Figure \ref{fig:lognorm_mult} as the blue curve.  We then provide the densities for the $1.1X$, $1.2X$, $1.3X$, $1.4X$ and $1.5X$ random variables (shown in red).  Hence, while these new densities are not specifically lognormal distributions, such changes in location in scale may be feasible practice.

\subsubsection{Using a three-point summary
$S_2 = \{q_1, m, q_3\}$}\label{sec:3_point_summary_est_S_2}

We now consider using a three-point summary in the form of the sample median, $m$, and interquartile range (IQR) $[q_1,\ q_3]$.  With limited information, more emphasis on distribution choice needs to be made, although there are some flexible three-parameter distributions that may be useful.  We will consider and explore one of the distributions below. 

The quantile-based skew logistic distribution (SLD), first proposed by \cite{gilchrist2000statistical}, arises from the linear combination of the exponential and reflected exponential distributions.  Since it was first introduced,  the quantile function has been modified \citep{van2015quantile}  and is given as
\begin{equation}
    Q(p)=\lambda +\eta\left[(1-\delta)\ln(p) - \delta\ln(1-p)\right], \ 0 \leq \delta \leq 1.
\end{equation}

When $\delta=0.5$, the skew logistic results, and $\delta=1$ or $-1$ provides the exponential distribution and reverse exponential distribution respectively.  The logistic distribution is similar in shape to the normal distribution and, for what we are trying to achieve here, this could be suitable for when the underlying distribution is normal.  Additionally, we found that the skew-logistic can provide a good approximation to the lognormal distribution, with this exception of those that are very highly skewed.

\begin{figure}[h!t]
    \centering
    \includegraphics[scale=0.4]{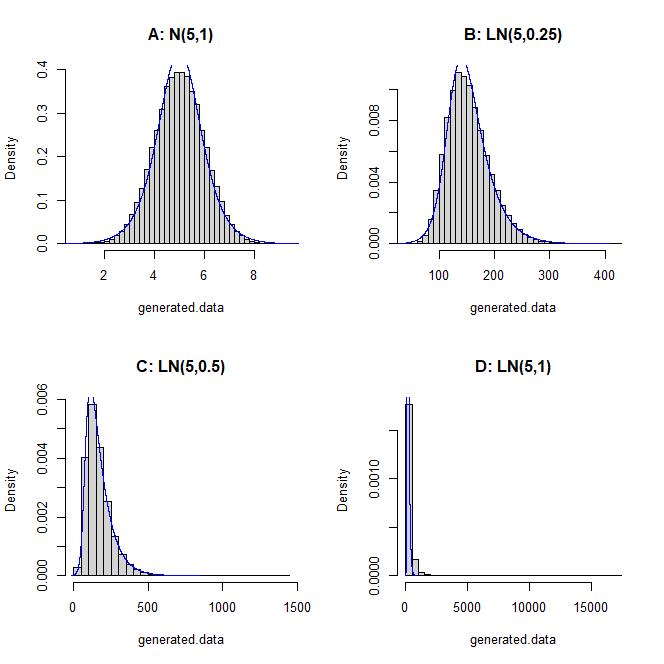}
    \caption{Histograms (probability density scale) of 100,000 data points randomly sampled from the $N(5,1)$, LN$(5, 0.25)$, LN$(5, 0.5)$ and LN$(5, 1)$ distributions.  The blue curve is the fitted skew logistic density with parameters estimated from the simulated data.}
    \label{fig:hist}
\end{figure}

In Figure \ref{fig:hist} we provide fitted skew logistic densities to 100,000 simulated data points from the $N(5,1)$, LN$(5, 0.25)$, LN$(5, 0.5)$ and LN$(5, 1)$ distributions where the estimates were obtained using the \texttt{sld} package \citep{sld}.  These were the distributions considered in Figure 3 of \cite{wan2014}.  As can be seen, the skew logistic does a reasonable job of fitting the normal data (Plot A) and the first two lognormal data sets (Plots B and C).  However, for the very highly skewed data in Plot D, the fit is poor.  In situations with such a high skew, it would be inadvisable to try and approximate the mean since it would not be representative of typical data.  It would be better to think of ways to use the median, and we look at one option later.  

Note that
\begin{equation*}
    \rho = \frac{Q(3/4) - Q(1/2)}{Q(1/2) - Q(1/4)}
\end{equation*}
only depends on $\delta$, and rearranging gives $\delta = \frac{\ln(3/2) - \ln(2)\rho}{\ln(3/4)(\rho + 1)}$.  This leads to an estimate of $\delta$,
\begin{equation}
    \widehat\delta = \frac{\ln(3/2) - \ln(2)\widehat{\rho}}{\ln(3/4)(\widehat{\rho} + 1)}\label{eq:delta_hat}
\end{equation}
where $\widehat{\rho}=(q_3-m)/(m-q_1)$.  We can then obtain an estimate of $\eta$ by re-arranging $Q(3/4)-Q(1/2)$, the width of the IQR, and using $q_1, q_3$ to obtain
\begin{equation}
    \widehat\eta = \frac{q_3-q_1}{\ln(3)}.\label{eq:eta_hat}
\end{equation}
Finally, we can use our estimates together with the sample median to estimate $\lambda$ which is given by
\begin{equation}
    \widehat\lambda=m + \widehat\eta\ln(2)(1-2\widehat\delta).\label{eq:lambda_hat}
\end{equation}

When we have two arms, treatment and control, with estimates of $\rho$ denoted $\widehat{\rho}_t$ and $\widehat{\rho}_c$ respectively, then under the assumption of a common $\delta$, a pooled estimate of $\delta$ is 
$$\widehat{\delta} = \frac{1}{\text{ln}(3/4)}\left\{w_t \frac{[\text{ln}(3/2)-\text{ln}(2)\widehat{\rho}_t]}{(\widehat{\rho}_t + 1)} +  w_c \frac{[\text{ln}(3/2)-\text{ln}(2)\widehat{\rho}_c]}{(\widehat{\rho}_c + 1)}\right\}.$$
We can then estimate $\eta$ and $\lambda$ for each of the treatment and control groups using the pooled estimate to $\delta$, \eqref{eq:eta_hat} and \eqref{eq:lambda_hat} with the within-arm quartile estimates.

\subsubsection{Using a three-point summary
$S_1 = \{a, m, b\}$}

In the case of having the minimum, $a$, and maximum, $b$, instead of $q_1$ and $q_3$, we use
\begin{equation*}
    \nu = \frac{Q(1-0.5/n) - Q(1/2)}{Q(1/2) - Q(0.5/n)}
\end{equation*}
which only depends on $\delta$.  As we did previously, we can solve this for $\delta$ and obtain the estimate
\begin{equation}\widehat{\delta} =  \frac{d_n - \ln(n)\widehat{\nu}}{[d_n-\text{ln}(n)](\widehat{\nu} + 1)}\label{eq:delta_hat2}
\end{equation} where $\widehat{\nu}=(b-m)/(m-a)$ and $d_n=\text{ln}[(2n-1)/n].$  As we did above for the $\{q_1,m,q_3\}$ case, an estimate for $\eta$ is
\begin{equation}
    \widehat\eta = \frac{b-a}{\ln(2n-1)}\label{eq:eta_hat2}
\end{equation} and the estimate to $\lambda$ is the same as in \eqref{eq:lambda_hat} with these new estimates of $\delta$ and $\eta$.

Similarly to the two arms and the three-point summary based on quartiles,  let $\widehat{\nu}_t$ and $\widehat{\nu}_c$ respectively denoted the estimators of $\nu$ for the treatment and control groups. A pooled estimate of $\delta$ is then
$$\widehat{\delta} = w_t \frac{d_{n_t} - \ln(n_t)\widehat{\nu}_t}{[d_{n_t}-\text{ln}(n_t)](\widehat{\nu}_t + 1)} +  w_c \frac{d_{n_c} - \ln(n_c)\widehat{\nu}_c}{[d_{n_c}-\text{ln}(n_c)](\widehat{\nu}_c + 1)}$$
where $n_t,n_c$ and $w_t,w_c$ are the sample sizes and weights for each arm as defined previously.

\subsection{Methods for visualizing distributions }\label{sec:visual}

Recently, \cite{shi2023detecting} provided methods to detect skewness from quantile summaries. They note that in the presence of skew, conversions to the mean and standard deviation can be unreliable, something for which we have attempted to rectify in this paper.  Another approach is to construct a visualisation of the approximated distributions for studies with only summary data to both (i) gain an understanding of the underlying distributions in general and the consequences for any resulting meta-analysis and (ii) to determine whether there is likely to be any differences in distributions between studies other than in location and scale.

\begin{table}[ht]
    \centering
    \begin{tabular}{lllllllll}
    \toprule
        Study & Group & Dist. & $n$ & $a$ & $q_1$ & $m$ & $q_3$ & $b$ \\ \midrule
        1 & Cont. & LN$(0.75, 1)$ & 75 & 0.23 & 1.36 & 2.47 & 4.23 & 23.37 \\
          & Trt.  & LN$(1.5, 0.35)$ & 100 & 2.29 & 3.64& 4.24& 5.28& 10.05\\
        2 & Cont. & GLD$(6, 0.1, 1.1, 1.8)$ & 50 & -2.03 & 0.87& 4.35& 8.81& 11.36 \\
          & Trt.  & GLD$(7, 0.15, 1.3, 1.8)$ & 100 & 1.89 & 4.27 &6.40 &8.93& 10.66\\
        3 & Cont. & N$(2.5, 1)$ & 150 & -0.09 & 2.05 &2.59 &3.26 & 5.15 \\
          & Trt.  & N$(7.5, 1.5)$ & 50 & 2.99& 6.49& 7.09& 8.45& 10.34\\
          \bottomrule
    \end{tabular}
    \caption{Five-point summary results from one simulation run for three studies, each with two arms (control and treatment) where Study 1 samples from lognormal distributions, Study 2 from the GLD and Study 3 from the normal, with varying parameters and sample sizes ($n$).}
    \label{tab:meta_dist}
\end{table}

Table \ref{tab:meta_dist} provides the five-point summaries for a single simulation run for three studies, each with two arms.  Within each study, we use the similar shapes of distributions but vary the parameters between control (Cont.) and treatment (Trt.).  In practice this would be akin to the treatment and control arms having some similarities having been drawn from the same population, but where a treatment effect results in differences between the two arms.  Study 1 samples from lognormal distributions with parameters $0.75, 1$ and $1.5, 0.35$.  For Study 2, we use the GLD distribution with parameter settings that result in a $U$-shaped distribution with a slight skew \citep[see, e.g., Table 7.7 of][]{gilchrist2000statistical}.  For Study 3 we sample from the $N(2.5, 1)$ and $N(7.5, 1.5)$ distributions.  Sample sizes are shown in the table, column $n$, as are the estimated five-point summaries.

\begin{figure}[h!t]
    \centering
    \includegraphics[scale=0.4]{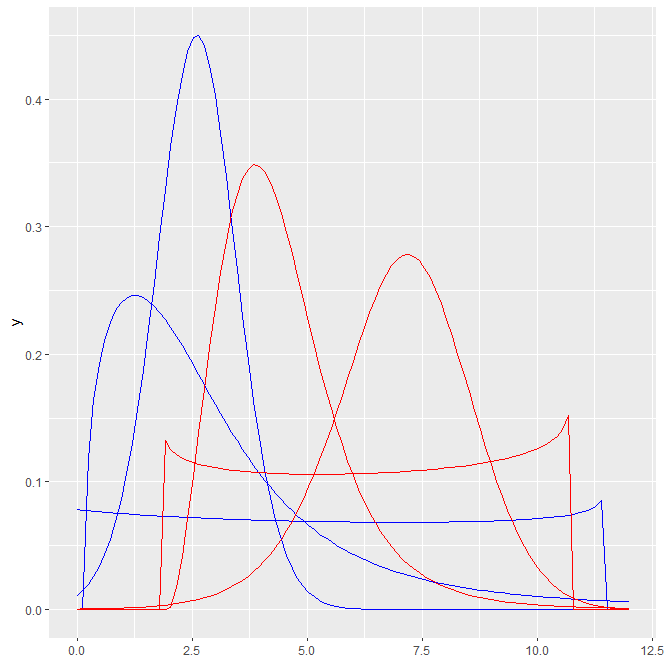}
    \caption{Plots of the densities for the three studies with data generated and summarised in Table \ref{tab:meta_dist}.  The control densities are displayed in blue and the treatment densities in red.  All densities have been approximated using the GLD and the five-point summaries.}
    \label{fig:meta_dist}
\end{figure}

In Figure \ref{fig:meta_dist} we display the approximated densities when using the GLD and the five-point summaries from the three simulated studies summarised in Table \ref{tab:meta_dist}.  The shapes are clearly distinguishable between the studies, and we can also gain an appreciation of differences at the study-level.  The two normal densities are evident (centered close to 2.5 and 7.5), as are the shapes of the two lognormal densities with obvious right-skew.  While skew for the GLD distribution is only slight, the densities are clearly non-normal, and have the $U$-shape as expected. Hence, in practice such an approach may be useful for visualising skew across several studies, but also for comparing shapes of underlying distributions between studies. Differences in shapes can result in heterogeneity and provide a starting point for further investigation. We have provided an example involving a real dataset in Section~\ref{subsec:dist_visualisation_eg}.

\subsection{Methods for the meta-analysis of quantiles and functions of quantiles}\label{sec:ma_quant}

Let $\widehat{x}_p$ be the sample quantile estimate of $x_p=Q(p)$.  An approximate variance of the quantile estimator is \citep{dasgupta2008asymptotic},
\begin{equation}
    \text{Var}(\widehat{x}_p)\approx \frac{p(1-p)}{n\left[f(x_p)\right]^2}\label{var_xp}
\end{equation}
where $f$ is the density function for the distribution from which the sample data was drawn.  Hence, using an estimate of the density, a Wald-type 95\% confidence interval for $x_p$ can be calculated by
\begin{equation}
    \widehat{x}_p \pm z_{0.975} \times \text{SE}\label{eq:CI_for_xp}
\end{equation}
where $z_{0.975}$ is the 97.5\% percentile of the standard normal distribution and SE is the standard error obtained by taking the square root of the estimated variance in \eqref{var_xp}.  It has been shown that intervals of this type can have very good coverage (i.e. close to the nominal $0.95$) for a wide variety of distributions and with only moderate sample sizes \citep{prendergast2016exploiting}.

Recently, methods for meta-analysing medians and quantiles have been developed.  For example, \cite{mcgrath2019one} use a median of medians approach where the median, or weighted median, of the medians reported from the $K$ independent studies is used.  Confidence intervals for the median are then quantile estimates from the reported medians. \cite{ozturk2020meta} considered inverse variance weights (IVW) meta-analysis of quantiles, including medians, for studies where a confidence interval for the quantile, or the variance of the quantile, has been reported.  

In another approach, in their PhD dissertation \cite{CharlesTiGray2020TaMo,graymeta} considered another method to IVW meta-analyse medians, differences and ratios of medians that only requires the median, IQR (or the range) and sample sizes.  To do so, the lognormal distribution, $\text{LN}(\mu,\sigma)$, density was used in the variance of the median, where the lognormal parameters were estimated from $m$ and IQR.  The $\mu$ parameters was taken to the log of the median, and this, together with the sample quartiles equated to the true IQR width $Q(3/4)-Q(1/4)$, or range equated to $Q(0.5/n)-Q(1-0.5/n)$, was used to obtain the estimate for $\sigma$.  Good confidence interval coverages for the meta-analysis of medians were obtained, even for underlying distributions that were not lognormal.  This was because, at the median, $1/f(m)$ with the lognormal distribution provided a very good approximation for many other distributions.  

\cite{mcgrath2020meta} considered a more general approach to the above, called \textit{quantile estimation (QE)}, for meta-analysis of the difference of medians for two-group comparisons.  Rather than use only the lognormal distribution, several distributions, namely normal, lognormal, Weibull and gamma distributions for different skew and shapes, are considered.  The density that minimizes a least squares criteria based on the observed and theoretical quantiles is chosen as the density to use in the variance estimate.  The density is chosen separately for each group and for each of the studies, and simulations shows that excellent coverages are obtained.  

\subsubsection{Intervals from summaries for quartiles}

The above highlight that an estimated density from the generalised lambda or skew logistic distribution could be useful as well, and also for other quantiles or functions of quantiles.

\begin{table}[ht]
\centering
\begin{tabular}{llllllll}
  \toprule
 $n$& Dist. & $Q_1$ 3-point & $Q_1$ 5-point  & $M$ 3-point & $M$ 5-point & $Q_3$ 3-point & $Q_3$ 5-point \\ 
  \midrule
 50 & N$(5,1)$ & 0.927 & 0.934 & 0.939 & 0.937 & 0.908 & 0.919 \\ 
  50 & LN$(5,0.25)$ & 0.937 & 0.939 & 0.937 & 0.926 & 0.887 & 0.926 \\ 
  50 & LN$(5,0.5)$ & 0.959 & 0.950 & 0.945 & 0.938 & 0.899 & 0.920 \\ 
  50 & LN$(5,1)$ & 0.984 & 0.966 & 0.944 & 0.945 & 0.857 & 0.9081 \\ 
  50 & EXP$(5)$ & 0.978 & 0.951 & 0.943 & 0.925 & 0.887 & 0.900 \\ 
  50 & BETA$(5,2)$ & 0.912 & 0.932 & 0.926 & 0.931 & 0.914 & 0.937 \\ 
  100 & N(5,1) & 0.937 & 0.954 & 0.939 & 0.943 & 0.936 & 0.945 \\ 
  100 & LN(5,0.25) & 0.923 & 0.945 & 0.923 & 0.926 & 0.918 & 0.922 \\ 
  100 & LN(5,0.5) & 0.931 & 0.964 & 0.944 & 0.946 & 0.920 & 0.927 \\ 
  100 & LN(5,1) & 0.966 & 0.956 & 0.962 & 0.956 & 0.894 & 0.907 \\ 
  100 & EXP(5) & 0.960 & 0.948 & 0.945 & 0.943 & 0.929 & 0.936 \\ 
  100 & BETA(5,2) & 0.927 & 0.936 & 0.939 & 0.949 & 0.938 & 0.945 \\ 
  200 & N(5,1) & 0.945 & 0.956 & 0.942 & 0.945 & 0.942 & 0.939 \\ 
  200 & LN(5,0.25) & 0.951 & 0.963 & 0.942 & 0.948 & 0.944 & 0.942 \\ 
  200 & LN(5,0.5) & 0.937 & 0.962 & 0.933 & 0.935 & 0.928 & 0.919 \\ 
  200 & LN(5,1) & 0.965 & 0.965 & 0.956 & 0.952 & 0.910 & 0.923 \\ 
  200 & EXP(5) & 0.972 & 0.957 & 0.954 & 0.952 & 0.941 & 0.944 \\ 
  200 & BETA(5,2) & 0.948 & 0.937 & 0.933 & 0.943 & 0.952 & 0.952 \\ 
   \bottomrule
\end{tabular}\caption{Simulated coverage probabilities for 95\% Wald-type confidence intervals for the first quartile $(Q_1)$, median $(M)$ and the third quartile $(Q_3)$ where 1000 datasets of size $n=50,100$ were drawn from normal, lognormal, exponential and beta distributions.  We consider both three-point and five-point summaries with skew logistic and GLD estimation of the densities respectively.}\label{tab:CP}
\end{table}

In Table \ref{tab:CP}, we provide empirical coverage probabilities for 95\% confidence intervals constructed using \eqref{eq:CI_for_xp} for the first quartile ($Q_1$), median $(M)$ and third quartile $(Q_3)$.  1000 datasets of size $n$ were sampled from several distributions, including the normal, lognormal (LN), exponential (EXP) and Beta (BETA) distributions, and only the three- or five-point summaries were used in the estimate of the interval.  The density for the variance estimate was obtained using either the three-point density estimator in Section \ref{sec:3_point_summary_est_S_2} or the five-point estimator of Section \ref{sec:5_point_summary_est}.  As can be seen, coverages are very close to the nominal 0.95 for interval estimators of $Q_1$ and $M$.  For $Q_3$ and for skewed distributions where the distribution mass around $Q_3$ is less dense, coverages can be lower although typically reasonable for the three-point summary.  Coverages for $Q_3$ improve when using the five-point summary.

\subsubsection{Intervals from summaries for comparing dispersion}

In a two-arm (e.g. treatment versus control) scenario, \cite{prendergast2016meta} conduct meta-analysis on ratios of sample variances to analyse differences in dispersion.  Similarly, we could do this with squared ratios of IQRs. \cite{arachchige2021interval} provide confidence intervals for ratios of log IQRs when data is available for two independent samples.  Let $q_{t1}$, $q_{c1}$ denote the estimated first quartiles for treatment and control, and similarly define $q_{t3}$ and $q_{c3}$ for the third quartile.  Further, let $r=(q_{t3}-q_{t1})^2/(q_{c3}-q_{c1})^2$ be the ratio of squared IQRs.  An approximate standard error for $r$ is \begin{equation}
\text{SE}_r =\left\{\frac{r}{n_1+n_2}\sum_{i=\in\{t,c\}}\frac{g_{i1}^2+g_{i3}^2-[g_{i1}+g_{i3}]^2}{w_i(q_{i3}-q_{i1})^2}\right\}^{1/2}\label{SE}
\end{equation}
where, e.g., $g_{t1}=1/\widehat{f}(q_{t1})$ and $\widehat{f}$ denotes the estimated density.  Then an approximate 95\% confidence interval for the true ratio of squared IQR widths is \citep{arachchige2021interval}
\begin{equation}
    \exp\left[\ln(r) \pm z_{0.975}\frac{1}{r}\times \text{SE}_r\right].
\end{equation}
As we did for the interval estimators of quartiles using three- and five-point summaries, we will now investigate whether we can obtain good interval estimates for the true ratio of squared IQRs using just the summary data.

\begin{table}[ht]
\centering
\begin{tabular}{lllllll}
  \toprule
  Dists. & $n=50$ & $n=50$ & $n75$ & $n=75$ & $n=100$ & $n=100$ \\ 
  & 3-point & 5-point & 3-point & 5-point & 3-point & 5-point \\
  \midrule
 N(5,1) vs N(5, 1.5) & 0.973 & 0.950 & 0.970 & 0.954 & 0.969 & 0.954  \\ 
   LN(5,0.25) vs LN(5, 0.5) & 0.966 & 0.949 & 0.954 & 0.935 & 0.954 & 0.942  \\ 
   LN(5,0.5) vs LN(5, 0.25) & 0.953 & 0.947 & 0.968 & 0.958 & 0.966 & 0.957  \\ 
   EXP(5) vs EXP(3) & 0.947 & 0.940 & 0.951 & 0.948 & 0.940 & 0.933  \\ 
   BETA(5, 2) vs BETA(4, 3) & 0.982 & 0.952 & 0.975 & 0.949 & 0.979 & 0.961  \\ 
   \bottomrule
\end{tabular}\caption{Simulated coverage probabilities for 95\% confidence intervals of single study squared ratio of IQR widths.  The distributions for each group are given in column Dists. with the first being for the numerator IQR and the second for the denominator.  The total number of trials was 1,000 for each scenario and for each choice of sample size $(n)$ with equal samples between groups for simplicity.}\label{table:cp_IQR}
\end{table}

Simulated coverage probabilities for three different sample sizes ($n_1=n_2=n$ for simplicity) are shown in Table \ref{table:cp_IQR}.  As can be seen, for a single study the coverages are very close to the nominal 95\% even for $n=50$.  Hence, when having only summary data available, intervals with good coverage properties can be obtained.

\section{Some further simulations and examples}\label{sec:sim_and_ex}

This section presents the results of some further simulations and examples to demonstrate the applications of the methods described in the previous section.

\subsection{Simulations comparing methods for estimating mean and standard deviations from quantile summaries}\label{subsec:sim_means}

In this section, we compare the proposed approaches with those from~\cite{luo2018, wan2014, shi2020, shi2023detecting} which are the default methods in the widely-used \texttt{metafor} \citep{metafor} and \texttt{meta} \citep{metapackage} packages in R. Let $\bar{x}_i$ and $\hat{x}_i$ denote the true sample mean and estimate of the sample mean of the $i$th simulated iteration respectively, for $i = 1, \ldots, 1000$. Similarly, let  $s_i$ and $\hat{s}_i$ denote the the true sample standard deviation and estimate of the sample standard deviation of the $i$th simulated iteration respectively. To compare the methods, we use the average relative error (ARE) of the sample estimators defined as, $$\text{ARE}(\hat{x}) = \frac{1}{1000} \sum_{i=1}^{1000} \frac{\hat{x}_i - \bar{x}_i}{\bar{x}_i}\;\;\text{and}\;\; \text{ARE}(\hat{s}) = \frac{1}{1000} \sum_{i=1}^{1000} \frac{\hat{s}_i - s_i}{s_i}.$$

For direct comparison with the published literature~\citep[e.g.,][]{hozo2005, luo2018}, we first present the following distributions : (i) normal distribution with mean $\mu=50$
 and standard deviation $\sigma=17$, (ii) log-normal distributions with location parameter $\mu=4$ and scale parameters $\sigma=0.3$, (iii) exponential distribution with rate parameter $\lambda=10$, and (iv) Weibull distribution with shape parameter $k=2$ and scale parameter $\lambda=35$. A total of $n=25,50,75, 100, \dots,1000$ observations were randomly generated from each distribution. For further evaluation, we additionally considered the following distributions, with the results presented in the Supplementary Information Section: normal distribution with mean $\mu=5$ and standard deviation $\sigma=1$,log-normal distributions with location parameter $\mu=5$ and scale parameters $\sigma=0.1, 0.25, 0.5, 1$ respectively, beta distribution with shape parameters $\alpha=9$ and $\beta=15$,  beta distribution with shape parameters $\alpha=2$ and $\beta=5$, and chi-square with $2$ degrees of freedom. The densities for all considered distributions are shown in Supplementary Figure S1.
 
For each distribution and each sample, we computed the sample mean, sample standard deviation and the quantile summaries for scenarios $S_1$, $S_2$, and $S_3$ described in Section~\ref{subsubsec:est_means_sd}. We then used the estimation approaches described in Section~\ref{sec:proposed} to obtain an estimate of the mean and standard deviation of each sample using only the corresponding quantile summary and calculate the quantities
 $\text{ARE}(\hat{x})$ and $\text{ARE}(\hat{s})$. 

 \begin{figure}[h!t]
    \centering   
    \includegraphics[width=1\textwidth]{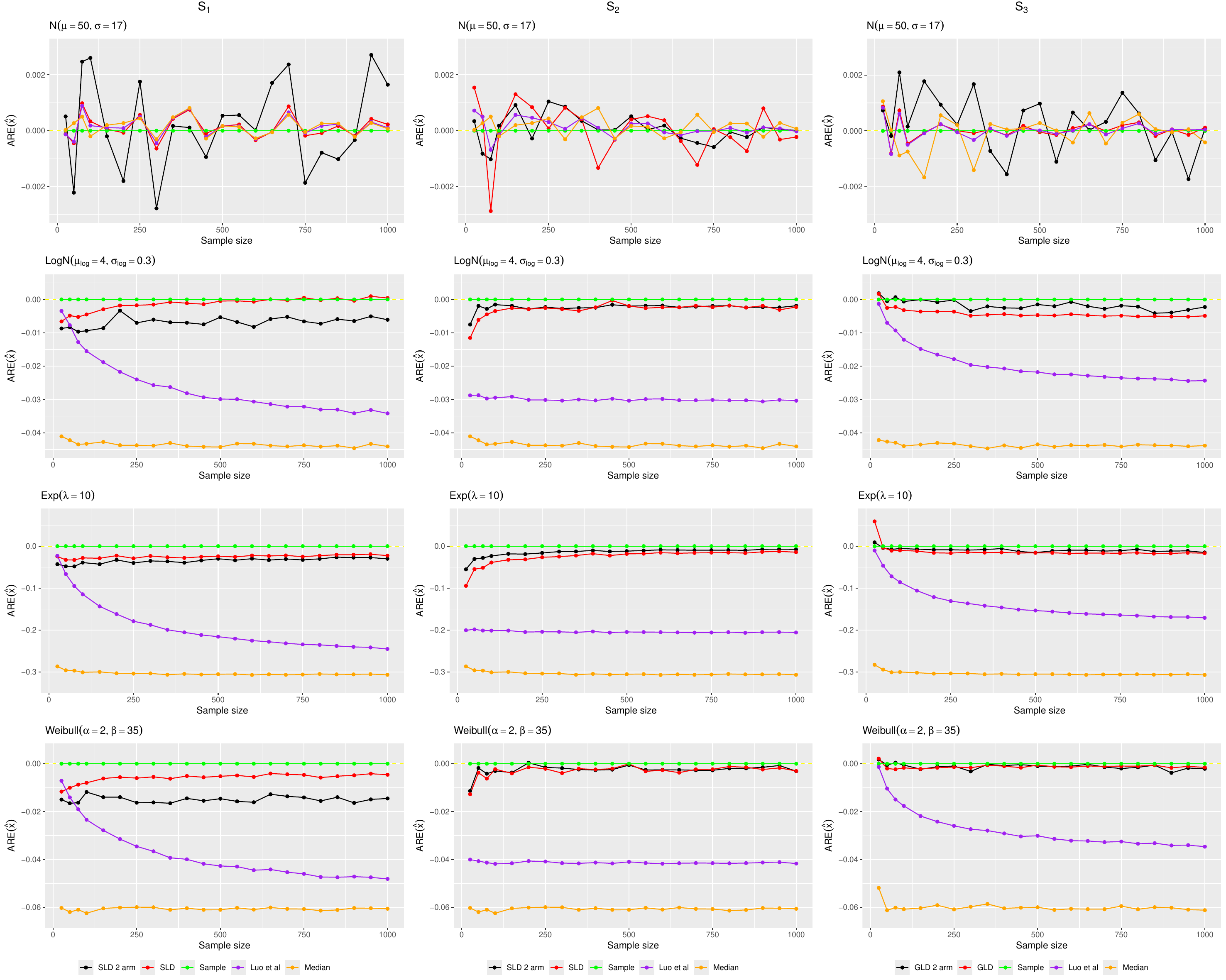}
    \caption{$\text{ARE}(\hat{x})$ for the proposed methods (skew logistic distribution (SLD), generalised lambda distribution (GLD), and their 2-arm variants), Luo et al.(2018) and `median' across all scenarios for the distributions (i-iv) mentioned in Section 3.1. The columns from left to right show the ARE for $S_1$, $S_2$, and $S_3$, respectively. The sample estimates of the mean are also shown. }
    \label{fig:ARE_means_all_scenarios}
\end{figure}

The $\text{ARE}(\hat{x})$ obtained for all scenarios is shown in Figure~\ref{fig:ARE_means_all_scenarios}. As researchers often use the median to replace mean in practice, the median is also shown in the plot along with sample estimates of the mean. The figure shows that the means estimated using generalised lambda and skew logistic distributions estimate the sample means as well as or better than the existing methods across all scenarios. An improvement in accuracy was not generally observed when two arm information was used in the estimation. As expected, replacing the mean by the median estimates the sample mean well when the underlying distribution is normal, but not when it is skewed. The results for further distributions are shown in Supplementary Figure S2 and are consistent with the above findings.

For completeness, we have presented the $\text{ARE}(\hat{s})$ obtained for all scenarios in Supplementary Figures S3 and S4. The figures show that the comparative performance of both the generalised lambda and skew logistic distributions when estimating the standard deviations is highly variable. The approaches proposed by \cite{wan2014}  and \cite{shi2020} show better performance across all scenarios except for the exponential distribution, log-normal distributions with a high standard deviation, and Chi-square distribution. 

\subsection{Distribution visualisation example}\label{subsec:dist_visualisation_eg}

In this section we consider the \texttt{dat\_age} dataset from the \texttt{metamedian} package~\citep{metamedian} which stems from a systematic review by \cite{katzenschlager2021can}.  The data consists of summary measures of age from 51 studies grouped into two groups following COVID-19 infection, survivors and non-survivors.  Much has been reported on age and its association with death due to COVID, and consequently it makes a useful illustrative example here.  However, many other measures are available in the \texttt{metamedian} package which may also be of interest.  We will focus on the 29 studies that report the quartiles and sample sizes only.

\begin{figure}[!ht]
    \centering
    \includegraphics[scale = 0.5]{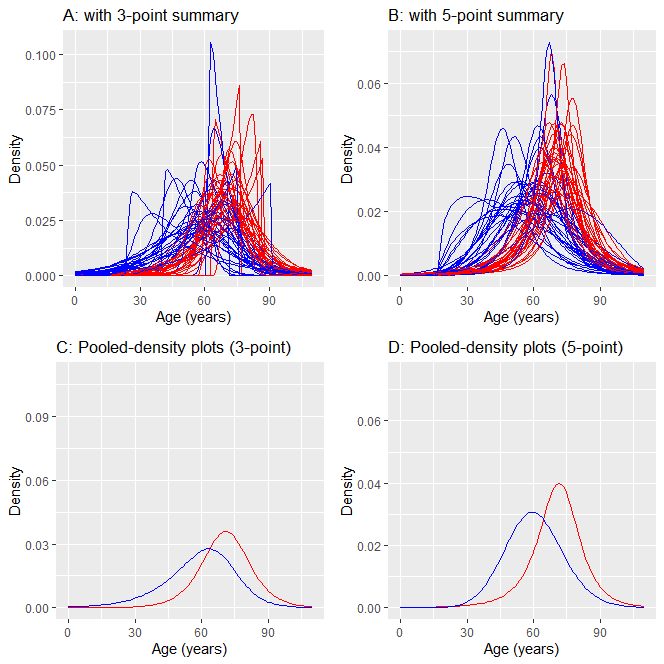}
    \caption{Density plots of age (years) of survivors (blue) and non-survivors (red) of COVID-19 infections from 29 studies.  Plot A includes the densities from the 3-point summaries using the skew logistic.  Plot B uses 5-point summaries where minimum age was taken to be 18, and maximum 110 years for all studies.  Plots C and D depict the pooled-estimated densities using the pooled parameter estimates.}
    \label{fig:age_plots}
\end{figure}

In Plot A of Figure \ref{fig:age_plots}, we plot the approximate densities for each of the 29 studies and for each group, survivors (blue) and non-survivors (red).  As expected, the densities for the non-survivor groups are typically shifted to the right, with older people being more susceptible to poor outcomes following infection.  We can also see that is often higher dispersion in the survivor group, with many elderly survivors in contrast the much fewer deaths among younger people.  Additionally, and analogous to a forest plot visualisation, heterogeneity in central location, dispersion and shape is evident.

Since the skew logistic is based only on the three quartiles, the fitted tails may not be very informative.  For the purposes of the visualisation of distributions, this may or may not be a problem.  We can see in Plot A that some density falls to the left of age zero in the survivors group, which of course is not possible in reality.  Further, most of the studies were on adult cohorts and so a density mass less than, say, 18 years is not accurate.  However, we may improve our approximations be using a five-point summary.  Many studies that report median and IQR will also report some evidence in support of plausible minima and maxima.  We have provide some examples selected from the 29 studies that make choosing these possible, and note that most studies, in total, had some information either minimum or maximum, or both.  

\begin{table}[ht]
    \centering
    \begin{tabular}{ll}
    \toprule
    Study & Details on minimum and/or maximum \\
    \midrule
    \cite{deng2020clinical}  & ``... age ranges of the death and recovered group were 33 to 94 and 22 to 81..." \\
    \cite{paranjpe2020clinical} & Includes frequencies: 73 in $[18,\ 30]$ and 82 who are $\geq 91$.\\
    \cite{zhou2020clinical} & ``...we included all adult inpatients ($\geq 18$ years old)..." \\
    \bottomrule
    \end{tabular}
    \caption{Examples of minimum and/or maximum details provide three of the studies in the meta-analysis of age for comparing COVID survivors and non-survivors.}
    \label{tab:example_min_max}
\end{table}

Some examples of additional information can be found in Table \ref{tab:example_min_max}.  In the \cite{deng2020clinical} study, the ranges in each group is clearly stated and so an exact five-point summary is achievable.  Frequencies for age ranges are provided in \cite{paranjpe2020clinical} indicating that all were age 18 years or greater, and some were older than 91 years.  In \cite{zhou2020clinical}, the inclusion criteria stipulates that only adults were reported, with adult being defined as being 18 years or older.  This type of information is typical, rather than uncommon and so we can include additional information to form five-point or approximate five-point summaries.  For the purpose of this illustration though, we have simply selected a minimum of 18 and a maximum of 110 for all studies.  In Plot B we can see the approximate densities using the GLD distribution and the five-point summaries.  Tail behavior is more realistic, although the conclusions from the visualisation remain the same as for Plot A.

\subsection{Quantile meta-analysis example}\label{subsec:quant_ma_eg}

Estimates and confidence intervals for comparing the quartiles and the IQR widths between the non-survivor and survivor groups are shown in Table \ref{tab:age_cis}.  Analyses were performed using the \texttt{metafor} package \citep{metafor} with a random effects model and using the REML estimator of the heterogeneity variance. All differences are highly statistically significant, and the pooled estimates and associated 95\% confidence intervals indicate a meaningful difference between the groups.  The estimates of the differences in quartiles verify what we could determine visually in Figure \ref{fig:age_plots}.  That is, the distribution of ages for the non-survivors is shifted to the right indicating that older people were more susceptible to death from COVID-19 complications.  Further, the mean ratio of squared IQR widths was just 0.48, 95\% CI $(0.37, 0.62)$ indicating the ages of the non-survivors were typically more concentrated.  This can explained by the fact that many of the elderly did survive COVID-19 infection, whereas fewer younger people died due to infection.

\begin{table}[ht]
    \centering
    \begin{tabular}{lcccc}
    \toprule
    Est.     &  $Q_{N1} - Q_{S1}$ & $M_{N} - M_{S}$ & $Q_{N3} - Q_{S3}$ & $\text{IQR}_N^2/\text{IQR}_S^2$ \\
    \midrule
    Pooled estimate & $15.349$ & $12.819$ & $10.793$ & $0.482$\\
    95\%  CI   &  $(12.160,\ 18.540)$ & $(10.203,\ 15.435)$ & $(8.715,\ 12.870)$ & $(0.374 0.620)$ \\
    $P-$value  & $<0.001$ & $<0.001$ & $<0.001$ & $<0.001$\\
    95\% PI & $(-0.350,\ 31.049)$ & $(-0.162,\ 25.800)$ & $(1.463,\ 20.122)$ & $(0.150,\ 1.55)$\\
    $I^2$ & $91.36\%$ & $90.05\%$ & $78.11\%$ & $92.34\%$ \\
    \bottomrule
    \end{tabular}
    \caption{Pooled estimates and associated 95\% confidence intervals and $P$-values following a meta-analysis of the difference in first quartiles $(Q_{N1}-Q_{S1})$, medians $(M_N-M_S)$, third quartiles $(Q_{N3}-Q_{S3})$ and also the ratio of squared IQRs.  Prediction intervals and $I^2$ as a measure of heterogeneity are also provided.}
    \label{tab:age_cis}
\end{table}

Based on $I^2$, there is a high degree of heterogeneity indicating that the observed variation in the quartiles and ratio of IQRs far exceeds what we would expect from estimator variability alone.  This agrees with our summary of heterogeneity from the visualisation of distributions in Figure \ref{fig:age_plots}.  95\% prediction intervals (PI) are a better way to demonstrate heterogeneity\citep{inthout2016plea}.  For example, the 95\% PI for the difference in first quartile tells us that approximately 95\% of future studies will have a difference in quartiles of between $-0.350$ and $31.049$.  Hence not all cohorts will have a first age quartile in the non-survivors group exceeding the survivors.  This was also true for the median, however, there was stronger evidence of a positive difference in the third quartile consistently across future studies.  The PI for the IQR ratio is very wide, telling as that while, on average there is less dispersion in ages for the non-survivors groups, there is an indication that this will not necessarily be the case across all studies.

Finally, we also consider the log-normal density approach of \cite{CharlesTiGray2020TaMo} and the QE method of \cite{mcgrath2020meta} for comparing the two medians.  The QE estimates were obtained using the \texttt{metamedian} package.  Both estimates return almost identical results, 12.820, 95\% CI $(10.202, 15.438)$ and 12.817, $(10.201, 15.434)$ respectively.  These are also almost identical to the results using the skew lambda distribution in Table \ref{tab:age_cis}, highlighting that many densities could provide a suitable choice when obtaining an approximate variance estimate for the median.

\section{Conclusions}\label{sec:conc}

Our contribution in this paper was two-fold. Firstly,
we proposed a novel-density based approach to  support a comprehensive meta-analysis when only the quantile summary measures are reported. We used flexible quantile-based distributions (namely, the generalised lambda and the skew logistic distributions) together with percentile matching to estimate the unknown parameters without making any prior assumptions about the underlying distribution. Our approach can be used to visualise unobserved distributions and to estimate unknown means using only the available quantile summary measures. When compared to widely-used methods for estimating the means in practice, our approach worked as well as or better than these existing methods across all scenarios considered.  

Secondly, we introduced quantile-based meta-analysis methods when a comparison of quantiles between groups themselves are of interest and found more suitable. We demonstrated these methods using simulations and real data. 

The R code for the methods proposed in this paper will be available via \url{https://github.com/metaanalysisR}.

\bibliographystyle{plainnat}
\bibliography{main}

\begin{thebibliography}{37}
\providecommand{\natexlab}[1]{#1}
\providecommand{\url}[1]{\texttt{#1}}
\expandafter\ifx\csname urlstyle\endcsname\relax
  \providecommand{\doi}[1]{doi: #1}\else
  \providecommand{\doi}{doi: \begingroup \urlstyle{rm}\Url}\fi

\bibitem[Anzures-Cabrera and Higgins(2010)]{anzures2010graphical}
Judith Anzures-Cabrera and Julian~PT Higgins.
\newblock Graphical displays for meta-analysis: an overview with suggestions
  for practice.
\newblock \emph{{Research Synthesis Methods}}, 1\penalty0 (1):\penalty0 66--80,
  2010.

\bibitem[Arachchige et~al.(2021)Arachchige, Cairns, and
  Prendergast]{arachchige2021interval}
Chandima~NPG Arachchige, Maxwell Cairns, and Luke~A Prendergast.
\newblock Interval estimators for ratios of independent quantiles and
  interquantile ranges.
\newblock \emph{{Communications in Statistics-Simulation and Computation}},
  50\penalty0 (12):\penalty0 3914--3930, 2021.

\bibitem[Bland(2015)]{bland2015}
Martin Bland.
\newblock Estimating mean and standard deviation from the sample size, three
  quartiles, minimum, and maximum.
\newblock \emph{International Journal of Statistics in Medical Research},
  4\penalty0 (1):\penalty0 57, 2015.

\bibitem[Borenstein et~al.(2011)Borenstein, Hedges, Higgins, and
  Rothstein]{borenstein2011introduction}
M.~Borenstein, L.V. Hedges, J.P.T. Higgins, and H.R. Rothstein.
\newblock \emph{Introduction to Meta-Analysis}.
\newblock Wiley, 2011.
\newblock ISBN 9781119964377.
\newblock URL \url{https://books.google.com.au/books?id=JQg9jdrq26wC}.

\bibitem[Borenstein et~al.(2017)Borenstein, Higgins, Hedges, and
  Rothstein]{borenstein2017basics}
Michael Borenstein, Julian~PT Higgins, Larry~V Hedges, and Hannah~R Rothstein.
\newblock Basics of meta-analysis: I2 is not an absolute measure of
  heterogeneity.
\newblock \emph{Research synthesis methods}, 8\penalty0 (1):\penalty0 5--18,
  2017.

\bibitem[DasGupta(2008)]{dasgupta2008asymptotic}
Anirban DasGupta.
\newblock \emph{Asymptotic theory of statistics and probability}, volume 180.
\newblock Springer, 2008.

\bibitem[Dedduwakumara et~al.(2019)Dedduwakumara, Prendergast, and
  Staudte]{dedduwakumara2019simple}
Dilanka~S Dedduwakumara, Luke~A Prendergast, and Robert~G Staudte.
\newblock A simple and efficient method for finding the closest generalized
  lambda distribution to a specific model.
\newblock \emph{Cogent Mathematics \& Statistics}, 6\penalty0 (1):\penalty0
  1602929, 2019.

\bibitem[Deng et~al.(2020)Deng, Liu, Liu, Fang, Shang, Zhou, Wang, Leng, Wei,
  Chen, et~al.]{deng2020clinical}
Yan Deng, Wei Liu, Kui Liu, Yuan-Yuan Fang, Jin Shang, Ling Zhou, Ke~Wang, Fan
  Leng, Shuang Wei, Lei Chen, et~al.
\newblock {Clinical characteristics of fatal and recovered cases of coronavirus
  disease 2019 in Wuhan, China: a retrospective study}.
\newblock \emph{{Chinese Medical Journal}}, 133\penalty0 (11):\penalty0
  1261--1267, 2020.

\bibitem[DerSimonian and Laird(1986)]{dersimonian1986meta}
Rebecca DerSimonian and Nan Laird.
\newblock Meta-analysis in clinical trials.
\newblock \emph{Controlled clinical trials}, 7\penalty0 (3):\penalty0 177--188,
  1986.

\bibitem[Freimer et~al.(1988)Freimer, Kollia, Mudholkar, and Lin]{Freimer1988}
Marshall Freimer, Georgia Kollia, Govind~S Mudholkar, and C~Thomas Lin.
\newblock A study of the generalized tukey lambda family.
\newblock \emph{Communications in Statistics-Theory and Methods}, 17\penalty0
  (10):\penalty0 3547--3567, 1988.

\bibitem[Gilchrist(2000)]{gilchrist2000statistical}
Warren Gilchrist.
\newblock \emph{Statistical modelling with quantile functions}.
\newblock Chapman and Hall/CRC, 2000.

\bibitem[Gray and Prendergast(2017)]{graymeta}
Charles Gray and Luke Prendergast.
\newblock Meta-analysis of medians.
\newblock In \emph{International Conference on Robust Statistics (ICORS): Book
  of Abstracts}, page~50, 2017.

\bibitem[Gray(2020)]{CharlesTiGray2020TaMo}
Charles~Ti Gray.
\newblock Towards a measure of code::proof: A toolchain walkthrough for
  computationally developing a statistical estimator, 2020.

\bibitem[Higgins and Thompson(2002)]{higgins2002quantifying}
Julian~PT Higgins and Simon~G Thompson.
\newblock Quantifying heterogeneity in a meta-analysis.
\newblock \emph{Statistics in medicine}, 21\penalty0 (11):\penalty0 1539--1558,
  2002.

\bibitem[Hozo et~al.(2005)Hozo, Djulbegovic, and Hozo]{hozo2005}
Stela~Pudar Hozo, Benjamin Djulbegovic, and Iztok Hozo.
\newblock Estimating the mean and variance from the median, range, and the size
  of a sample.
\newblock \emph{BMC medical research methodology}, 5:\penalty0 1--10, 2005.

\bibitem[IntHout et~al.(2016)IntHout, Ioannidis, Rovers, and
  Goeman]{inthout2016plea}
Joanna IntHout, John~PA Ioannidis, Maroeska~M Rovers, and Jelle~J Goeman.
\newblock Plea for routinely presenting prediction intervals in meta-analysis.
\newblock \emph{BMJ open}, 6\penalty0 (7):\penalty0 e010247, 2016.

\bibitem[Karian and Dudewicz(1999)]{karian1999fitting}
Zaven~A Karian and Edward~J Dudewicz.
\newblock Fitting the generalized lambda distribution to data: a method based
  on percentiles.
\newblock \emph{{Communications in Statistics-Simulation and Computation}},
  28\penalty0 (3):\penalty0 793--819, 1999.

\bibitem[Katzenschlager et~al.(2021)Katzenschlager, Zimmer, Gottschalk,
  Grafeneder, Schmitz, Kraker, Ganslmeier, Muth, Seitel, Maier-Hein,
  et~al.]{katzenschlager2021can}
Stephan Katzenschlager, Alexandra~J Zimmer, Claudius Gottschalk, J{\"u}rgen
  Grafeneder, Stephani Schmitz, Sara Kraker, Marlene Ganslmeier, Amelie Muth,
  Alexander Seitel, Lena Maier-Hein, et~al.
\newblock Can we predict the severe course of covid-19-a systematic review and
  meta-analysis of indicators of clinical outcome?
\newblock \emph{PLoS One}, 16\penalty0 (7):\penalty0 e0255154, 2021.

\bibitem[King and {van Staden}(2022)]{sld}
Robert King and Paul {van Staden}.
\newblock \emph{sld: Estimation and Use of the Quantile-Based Skew Logistic
  Distribution}, 2022.
\newblock URL \url{https://CRAN.R-project.org/package=sld}.
\newblock R package version 1.0.1.

\bibitem[Luo et~al.(2018)Luo, Wan, Liu, and Tong]{luo2018}
Dehui Luo, Xiang Wan, Jiming Liu, and Tiejun Tong.
\newblock Optimally estimating the sample mean from the sample size, median,
  mid-range, and/or mid-quartile range.
\newblock \emph{Statistical methods in medical research}, 27\penalty0
  (6):\penalty0 1785--1805, 2018.

\bibitem[McGrath et~al.(2019)McGrath, Zhao, Qin, Steele, and
  Benedetti]{mcgrath2019one}
Sean McGrath, XiaoFei Zhao, Zhi~Zhen Qin, Russell Steele, and Andrea Benedetti.
\newblock One-sample aggregate data meta-analysis of medians.
\newblock \emph{Statistics in medicine}, 38\penalty0 (6):\penalty0 969--984,
  2019.

\bibitem[McGrath et~al.(2020{\natexlab{a}})McGrath, Sohn, Steele, and
  Benedetti]{mcgrath2020meta}
Sean McGrath, Hojoon Sohn, Russell Steele, and Andrea Benedetti.
\newblock Meta-analysis of the difference of medians.
\newblock \emph{Biometrical Journal}, 62\penalty0 (1):\penalty0 69--98,
  2020{\natexlab{a}}.

\bibitem[McGrath et~al.(2020{\natexlab{b}})McGrath, Zhao, Steele, Thombs,
  Benedetti, and Collaboration]{mcgrath2020}
Sean McGrath, XiaoFei Zhao, Russell Steele, Brett~D Thombs, Andrea Benedetti,
  and DEPRESsion Screening Data~(DEPRESSD) Collaboration.
\newblock Estimating the sample mean and standard deviation from commonly
  reported quantiles in meta-analysis.
\newblock \emph{Statistical methods in medical research}, 29\penalty0
  (9):\penalty0 2520--2537, 2020{\natexlab{b}}.

\bibitem[McGrath et~al.(2023)McGrath, Zhao, Katzenschlager, Ozturk, Steele, and
  Benedetti]{metamedian}
Sean McGrath, XiaoFei Zhao, Stephan Katzenschlager, Omer Ozturk, Russell
  Steele, and Andrea Benedetti.
\newblock \emph{metamedian: Meta-Analysis of Medians}, 2023.
\newblock URL \url{https://CRAN.R-project.org/package=metamedian}.
\newblock R package version 1.1.1.

\bibitem[Ozturk and Balakrishnan(2020)]{ozturk2020meta}
Omer Ozturk and Narayanaswamy Balakrishnan.
\newblock Meta-analysis of quantile intervals from different studies with an
  application to a pulmonary tuberculosis data.
\newblock \emph{Statistics in Medicine}, 39\penalty0 (29):\penalty0 4519--4537,
  2020.

\bibitem[Paranjpe et~al.(2020)Paranjpe, Russak, De~Freitas, Lala, Miotto, Vaid,
  Johnson, Danieletto, Golden, Meyer, et~al.]{paranjpe2020clinical}
Ishan Paranjpe, Adam~J Russak, Jessica~K De~Freitas, Anuradha Lala, Riccardo
  Miotto, Akhil Vaid, Kipp~W Johnson, Matteo Danieletto, Eddye Golden, Dara
  Meyer, et~al.
\newblock Clinical characteristics of hospitalized covid-19 patients in new
  york city.
\newblock \emph{{MedRxiv}}, pages 2020--04, 2020.

\bibitem[Prendergast and
  Staudte(2016{\natexlab{a}})]{prendergast2016exploiting}
Luke~A Prendergast and Robert~G Staudte.
\newblock Exploiting the quantile optimality ratio in finding confidence
  intervals for quantiles.
\newblock \emph{Stat}, 5\penalty0 (1):\penalty0 70--81, 2016{\natexlab{a}}.

\bibitem[Prendergast and Staudte(2016{\natexlab{b}})]{prendergast2016meta}
Luke~A Prendergast and Robert~G Staudte.
\newblock Meta-analysis of ratios of sample variances.
\newblock \emph{Statistics in {M}edicine}, 35\penalty0 (11):\penalty0
  1780--1799, 2016{\natexlab{b}}.

\bibitem[{R Core Team}(2023)]{R}
{R Core Team}.
\newblock \emph{R: A Language and Environment for Statistical Computing}.
\newblock R Foundation for Statistical Computing, Vienna, Austria, 2023.
\newblock URL \url{https://www.R-project.org/}.

\bibitem[Schwarzer et~al.(2007)]{metapackage}
Guido Schwarzer et~al.
\newblock meta: An r package for meta-analysis.
\newblock \emph{R news}, 7\penalty0 (3):\penalty0 40--45, 2007.

\bibitem[Shi et~al.(2020)Shi, Luo, Weng, Zeng, Lin, Chu, and Tong]{shi2020}
Jiandong Shi, Dehui Luo, Hong Weng, Xian-Tao Zeng, Lu~Lin, Haitao Chu, and
  Tiejun Tong.
\newblock Optimally estimating the sample standard deviation from the
  five-number summary.
\newblock \emph{Research synthesis methods}, 11\penalty0 (5):\penalty0
  641--654, 2020.

\bibitem[Shi et~al.(2023)Shi, Luo, Wan, Liu, Liu, Bian, and
  Tong]{shi2023detecting}
Jiandong Shi, Dehui Luo, Xiang Wan, Yue Liu, Jiming Liu, Zhaoxiang Bian, and
  Tiejun Tong.
\newblock Detecting the skewness of data from the five-number summary and its
  application in meta-analysis.
\newblock \emph{{Statistical Methods in Medical Research}}, 32\penalty0
  (7):\penalty0 1338--1360, 2023.

\bibitem[van Staden and King(2015)]{van2015quantile}
P.~J. van Staden and R.~A.~R. King.
\newblock The quantile-based skew logistic distribution.
\newblock \emph{{Statistics \& Probability Letters}}, 96:\penalty0 109--116,
  2015.

\bibitem[Viechtbauer(2005)]{viechtbauer2005bias}
Wolfgang Viechtbauer.
\newblock Bias and efficiency of meta-analytic variance estimators in the
  random-effects model.
\newblock \emph{Journal of Educational and Behavioral Statistics}, 30\penalty0
  (3):\penalty0 261--293, 2005.

\bibitem[Viechtbauer(2010)]{metafor}
Wolfgang Viechtbauer.
\newblock Conducting meta-analyses in {R} with the {metafor} package.
\newblock \emph{Journal of Statistical Software}, 36\penalty0 (3):\penalty0
  1--48, 2010.
\newblock \doi{10.18637/jss.v036.i03}.

\bibitem[Wan et~al.(2014)Wan, Wang, Liu, and Tong]{wan2014}
Xiang Wan, Wenqian Wang, Jiming Liu, and Tiejun Tong.
\newblock Estimating the sample mean and standard deviation from the sample
  size, median, range and/or interquartile range.
\newblock \emph{BMC medical research methodology}, 14:\penalty0 1--13, 2014.

\bibitem[Zhou et~al.(2020)Zhou, Yu, Du, Fan, Liu, Liu, Xiang, Wang, Song, Gu,
  et~al.]{zhou2020clinical}
Fei Zhou, Ting Yu, Ronghui Du, Guohui Fan, Ying Liu, Zhibo Liu, Jie Xiang,
  Yeming Wang, Bin Song, Xiaoying Gu, et~al.
\newblock Clinical course and risk factors for mortality of adult inpatients
  with covid-19 in wuhan, china: a retrospective cohort study.
\newblock \emph{The lancet}, 395\penalty0 (10229):\penalty0 1054--1062, 2020.

\end{thebibliography}

\section*{Supplementary Information}

\begin{figure*}[ht]
    \centering
    \makebox[\textwidth]{\includegraphics[width=\textwidth,height=\textheight,keepaspectratio]{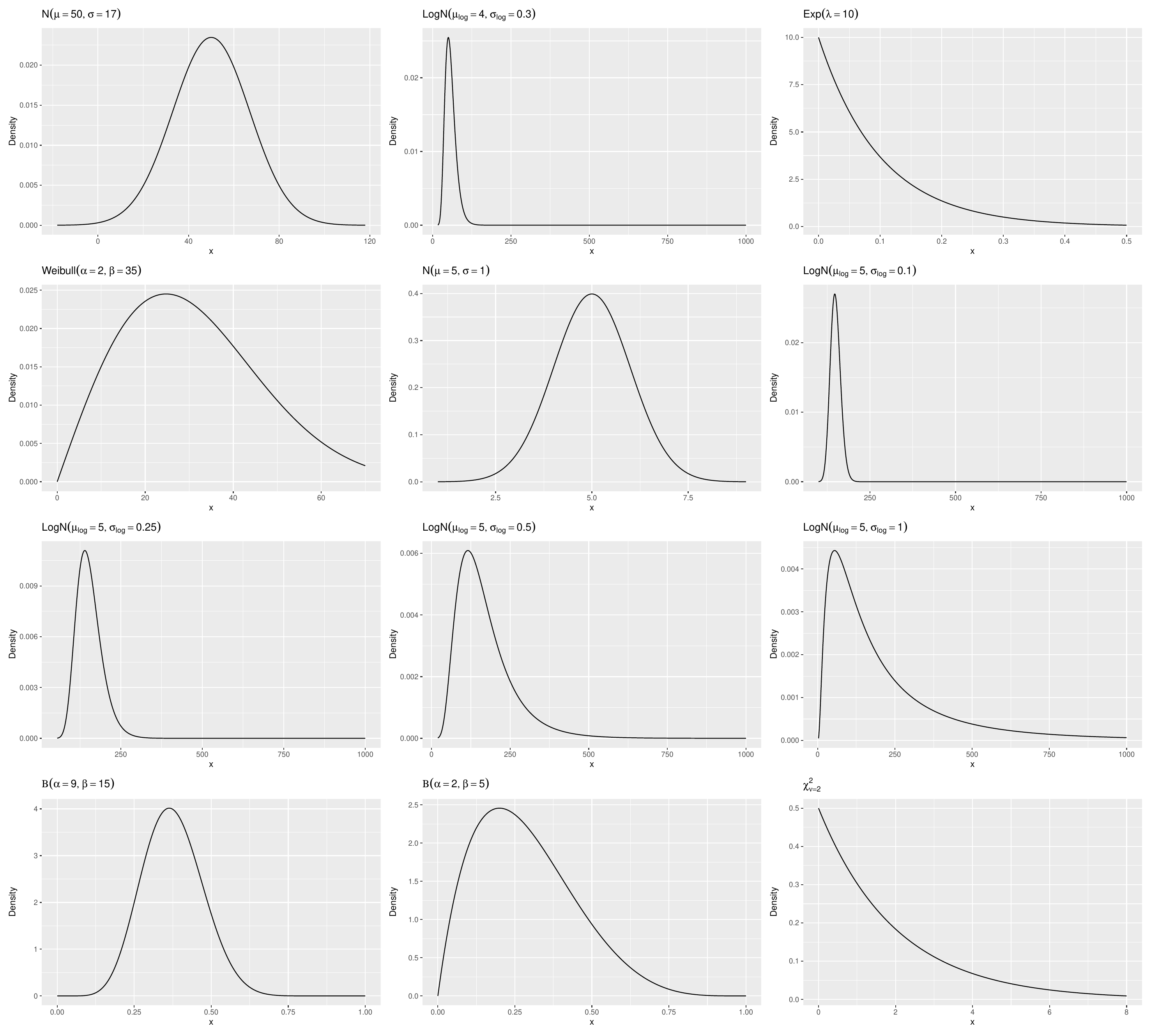}}
    \caption*{Supplementary Figure S1: Densities of the distributions considered in Section 3.1}
    \label{fig:dists}
\end{figure*}

 \begin{figure*}
    \centering       \includegraphics[width=1\textwidth,height=0.9\textheight,keepaspectratio]{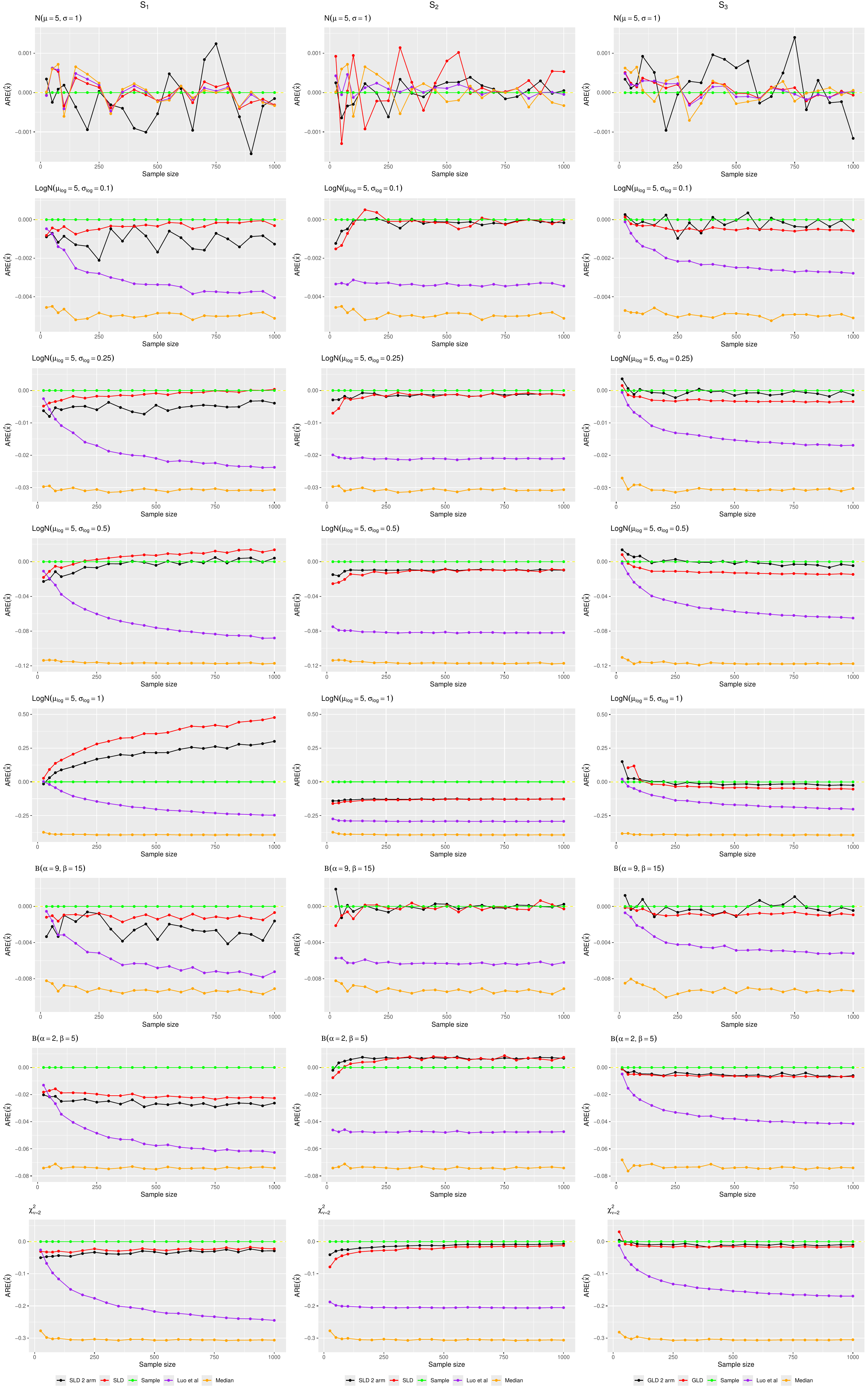}
    \caption*{Supplementary Figure S2:
    $\text{ARE}(\hat{x})$ for the proposed methods (skew logistic distribution (SLD), generalised lambda distribution (GLD), and their 2-arm variants), Luo et al.(2018) and `median' across all scenarios for the additional distributions mentioned in Section 3.1. The columns from left to right show the ARE for $S_1$, $S_2$, and $S_3$, respectively. The sample estimates of the mean are also shown.}    \label{fig:ARE_means_all_scenarios_additional}
\end{figure*}

 \begin{figure*}[h!t]
    \centering   
    \includegraphics[width=1\textwidth]{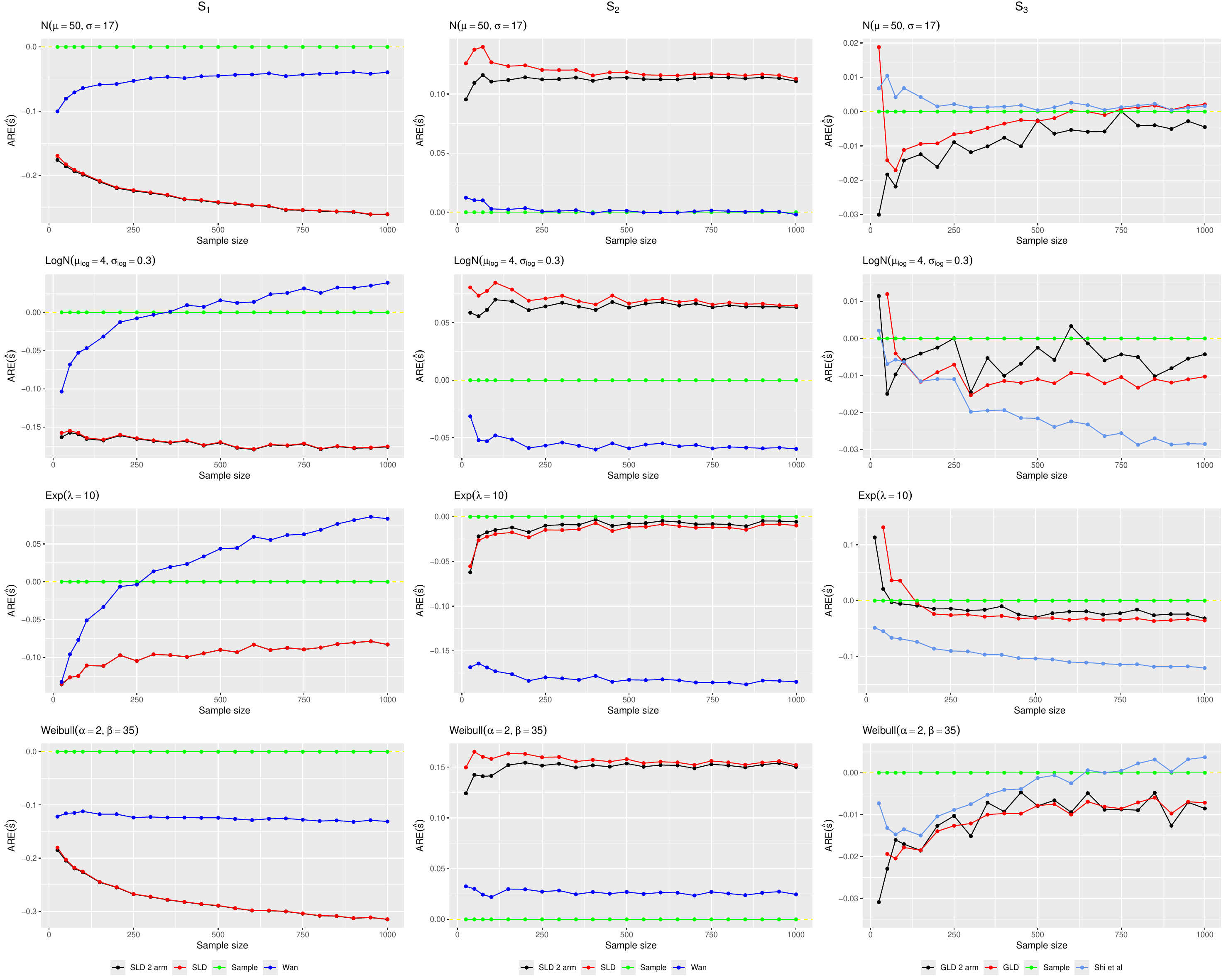}
    \caption*{Supplementary Figure S3:$\text{ARE}(\hat{s})$ for the proposed methods (skew logistic distribution (SLD), generalised lambda distribution (GLD), and their 2-arm variants), and the existing methods (Wan et al./Shi et al) across all scenarios for the distributions (i-iv) mentioned in Section 3.1. The columns from left to right show the ARE for $S_1$, $S_2$, and $S_3$, respectively. The y-axes were allowed to vary across scenarios.} \label{fig:ARE_sds_all_scenarios}
\end{figure*}

 \begin{figure*}
    \centering       \includegraphics[width=1\textwidth,height=0.9\textheight,keepaspectratio]{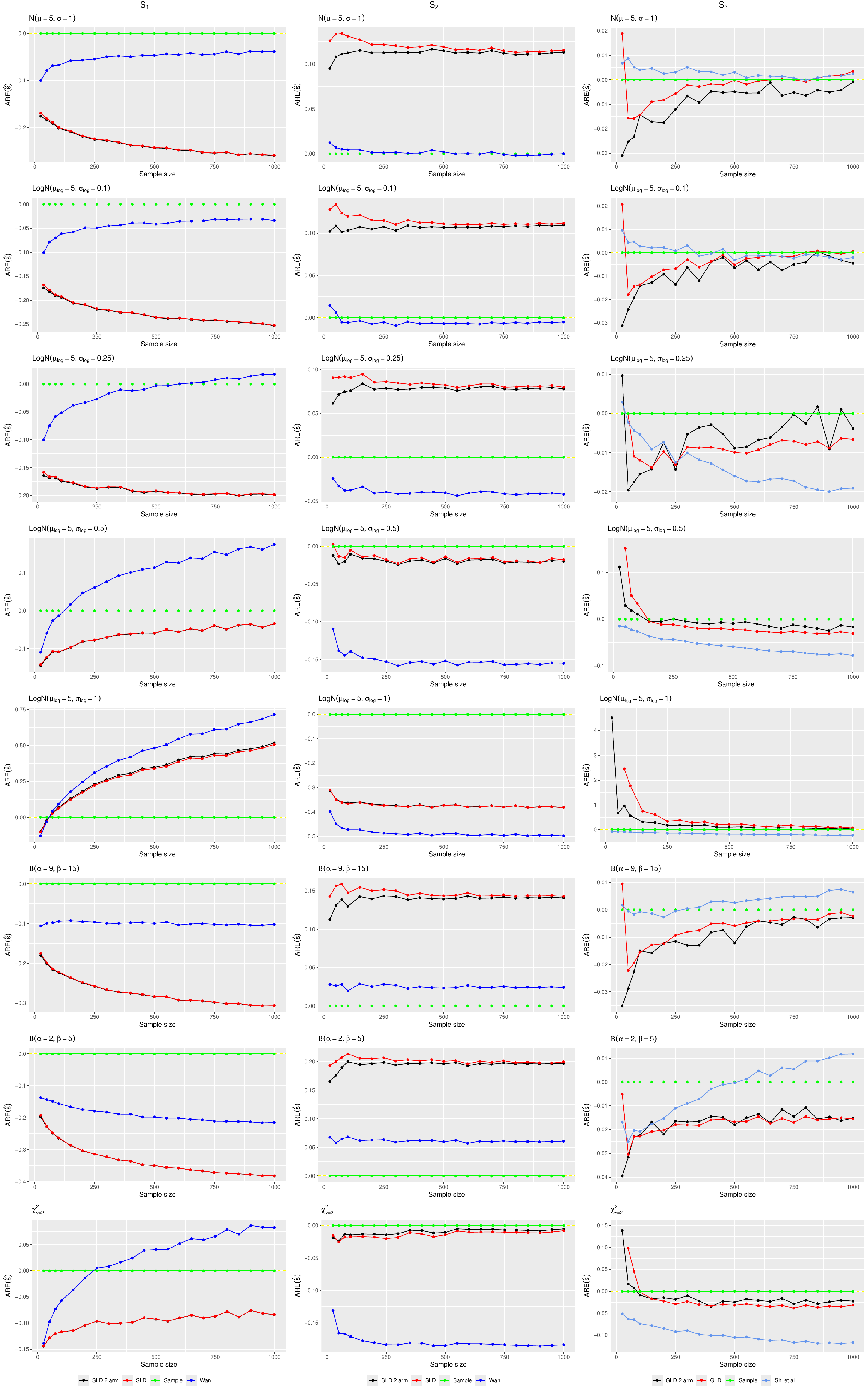}
    \caption*{Supplementary Figure S5:$\text{ARE}(\hat{s})$ for the proposed methods (skew logistic distribution (SLD), generalised lambda distribution (GLD), and their 2-arm variants), and the existing methods (Wan et al./Shi et al) across all scenarios for the additional distributions mentioned in Section 3.1. The columns from left to right show the ARE for $S_1$, $S_2$, and $S_3$, respectively. The y-axes were allowed to vary across scenarios.}   \label{fig:ARE_sds_all_scenarios_additional}
\end{figure*}

\end{document}